\documentclass{article}

\usepackage{microtype}
\usepackage{graphicx}
\usepackage{subfigure}
\usepackage{multirow}
\usepackage{booktabs} 
\usepackage{tablefootnote}
\usepackage{caption}
\usepackage{hyperref}
\usepackage{url}
\usepackage{breakurl}
\usepackage{xurl}  
\usepackage{bm} 

\usepackage{hyperref}

\usepackage[accepted]{icml2023}

\usepackage{amsmath}
\usepackage{amssymb}
\usepackage{mathtools}
\usepackage{amsthm}

\usepackage[capitalize,noabbrev]{cleveref}

\theoremstyle{plain}

\theoremstyle{definition}

\theoremstyle{remark}

\usepackage[textsize=tiny]{todonotes}

\icmltitlerunning{Quantifying the Causal Effect of Financial Literacy Courses on Financial Health}

\begin{document}

\twocolumn[
\icmltitle{Quantifying the Causal Effect of Financial Literacy Courses \\ on Financial Health}



\icmlsetsymbol{equal}{*}

\begin{icmlauthorlist}
\icmlauthor{Arnav Gangal}{equal,yyy}
\icmlauthor{Charles Shaviro}{equal,yyy}
\icmlauthor{Daniel Frees}{equal,yyy}
\end{icmlauthorlist}

\icmlaffiliation{yyy}{Stanford University}

\icmlcorrespondingauthor{Arnav Gangal}{agangal@stanford.edu}
\icmlcorrespondingauthor{Charles Shaviro}{cshaviro@stanford.edu}
\icmlcorrespondingauthor{Daniel Frees}{dfrees@stanford.edu}

\icmlkeywords{Causal Inference, Finance, Census, High School, Education, Propensity Score, IPW, AIPW, Sensitivity Analysis}


\vskip 0.3in
]



\printAffiliationsAndNotice{\icmlEqualContribution} 

\begin{abstract}
    In this study, we investigate the causal effect of financial literacy education on a composite financial health score constructed from 17 self-reported financial health and distress metrics ranging from spending habits to confidence in ability to repay debt to day-to-day financial skill. Leveraging data from the 2021 National Financial Capability Study, we find a significant and positive average treatment effect of financial literacy education on financial health. To test the robustness of this effect, we utilize a variety of causal estimators (Generalized Lin's estimator, 1:1 propensity matching, IPW, and AIPW) and conduct sensitivity analysis using alternate health outcome scoring and varying caliper strengths. Our results are robust to these changes. The robust positive effect of financial literacy education on financial health found here motivates financial education for all individuals and holds implications for policymakers seeking to address the worsening debt problem in the U.S, though the relatively small magnitude of effect demands further research by experts in the domain of financial health.\end{abstract}

\section{Introduction}\label{sec:introduction}


Consumer debt is widely understood to be a malignant and growing problem in the U.S. Measures of consumer debt are as high as they have ever been: Americans’ credit card debt recently crested over \$1 trillion dollars for the first time, student loan debt now exceeds \$1.7 trillion, and mortgage debt is over \$20 trillion \cite{creditcarddebt, studentloandebt, mortgagedebt}. For young people in particular, debt is a severe problem; almost one in five age 18–24 Americans with a credit record have debt in collections \cite{youthdebturbaninstitute}. 

An issue as pervasive and complex as mounting consumer debt fundamentally has many angles from which policymakers can attempt to address it. In this paper, we consider one popular method for reversing its growth: efforts to increase financial literacy. In an increasingly complex world where significant economic decisions are perpetually a click away in one's pocket, it stands to reason that financial decision-making is as complicated as its ever been. Can financial literacy education then help people make sounder financial decisions? Many seem to think so. There are consistent calls for greater financial education for Americans \cite{wapo_editorial_board} \cite{brookings} \cite{stanfordsiepr} and a  number of states have enacted financial education requirements for high school students. In Tennessee, for example, completing a personal finance class has been a requirement for graduating high school since 2013 \cite{tnFinancialEducation}. In several more states, similar bills have already been voted through or taken effect \cite{ramseysolutionsWhichStates}. But do financial literacy classes really encourage better decisions and lead to better financial health?

In this analysis, we use the National Financial Capability Survey (NFCS) to assess the causal relationship between financial education and financial health outcomes. The primary analysis assesses the causal effect of financial literacy education on financial health outcomes, while the secondary analysis focuses more specifically on the effect of high-school based financial literacy education. Through this assessment, we hope to inform policymakers about financial education policies' ability to positive impact Americans' finances.

\section{Related Work}\label{sec:related_work}
The financial well-being of young Americans, and of college students in particular, has been studied extensively. Much existing research focuses on financial health in the context of student loans, and on the impact that debt repayment can have on financial independence. A study conducted by \cite{fan2019financial} used the 2015 NFCS to examine the association of financial education and financial socialization with student loan repayments, but did not perform any causal analysis. Instead, it focused on associations between correct responses to financial knowledge questions, and whether individuals were on time with their student loan repayments. 

Another study of the NFCS found that individuals with outstanding student loans were more likely to have other substantial debt obligations, such as credit card debt or car repayments \cite{fry2012record}. These findings were supported by more recent research \citep{lusardi2023importance}, which indicated that areas with poor average financial literacy had higher wealth inequality, and that financial literacy metrics were heavily imbalanced across demographic categories such as race and age. 

We seek to improve upon these studies by analyzing the impact of financial education across a more comprehensive outcome measure of financial health, and by employing robust causal frameworks to isolate the effect of financial education.


\section{Data Features and Preprocessing}\label{sec:data}

\subsection{Data Source}

In order to assess the effect of financial education on financial health outcomes, we used data downloaded from the National Financial Capability Study (NFCS) \cite{finrafoundation}, commissioned by the FINRA Investor Education Foundation. The NFCS surveys a representative sample of approximately 500 people from each state, asking questions about their demographic background and level of education, and assessing their financial situation in terms of credit card debt, retirement savings, mortgage payments, and more. The survey began in 2009, and repeats every three years, with most recent data from 2021. For our study, we downloaded the 2021 archive of NFCS data. We opted not to include earlier data because many states implemented financial literacy laws for high school within the last ten years, and the 2021 dataset is recent enough to reflect those laws \cite{urban2020effects}. 

Two datasets were formed from the NFCS data. The primary dataset included participants who took a financial literacy course (treatment) in high-school, college, at work, or in the military, as well as participants who were certain they had never taken such a course (controls). Our second subset of data, designed for specifically analyzing the effect of high-school (HS) financial literacy education on financial health outcomes, includes participants who took a high-school financial literacy course, and participants who were sure they had not.

\subsection{Data Cleaning}

All data cleaning and feature engineering was performed using Python's \texttt{pandas} library \cite{pandas}. The associated data fact sheet for our downloaded 2021 NFCS data was used to derive a column map and rename columns to more appropriate and descriptive names. All strings were stripped of starting and trailing whitespace, string representations of integers were converted to integers, and NaN values were removed. 

$17$ columns were identified as critical markers of financial health (see \autoref{sec:health_markers}). Given the inconsistency of the original answer choices for each survey question corresponding to these markers, results were scaled so that answer values ranged from $0$ to $9$ for all markers, with $0$ indicating poor relative financial health and $9$ indicating great relative financial health. In cases where the respondent selected $98$ (Don't know) or $99/999$ (Prefer not to say) for a given question, we imputed a score of $4.5$, right in the middle of the value range. 

\subsection{Covariates}

Covariates were selected such that we could better control for the effects of financial literacy education on long-term financial health outcomes with fewer confounding effects. As such, we chose $9$ variables which we determined to be likely to impact financial health, but which were not financial health markers, and therefore should not be designated as outcome variables (see \autoref{sec:cov_list}).

\begin{figure}[!htb]
    \vskip 0.1in
    \begin{center}
        \caption{Slightly imbalanced age groups (primary dataset). \vspace{2mm}}
        \includegraphics[width=0.5\columnwidth]{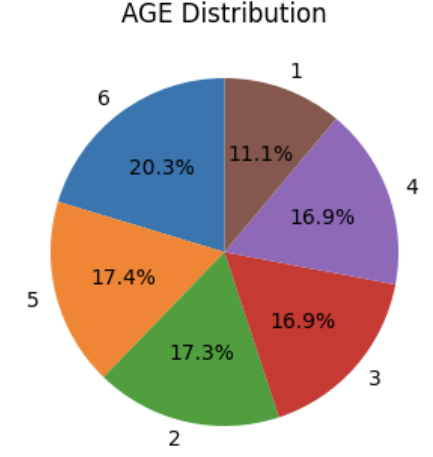}
        \label{fig:age}
    \end{center}
    \tiny Buckets and their corresponding age ranges: \\ 1) 18-24, 2) 25-34, 3) 35-44, 4) 45-54, 5) 55-64, and 6) 65+.
    \vskip -0.1in
\end{figure}

Covariates were distributed similarly in both our primary dataset and our secondary HS analysis dataset. Most participants had no children (answer option $6$) or no financially dependent children (answer option $5$). Of those that had children, most had $1$, followed by $2$, followed by far fewer with $3$ or more. Approximately $20\%$ of participants in both datasets were laid off due to COVID-19. Gender was very balanced in both datasets, with approximately $53\%$ male participants. 
For the HS dataset, age distribution was well-balanced across the six age buckets. Ages were somewhat imbalanced on the primary analysis dataset (\autoref{fig:age}), with more representation in the senior age group ($6$) as compared with the young adult age group ($1$). 

\begin{figure}[!htb]
    \vskip 0.1in
    \begin{center}
        \caption{CA and OR with increased state representation (primary dataset).}
        \includegraphics[width=0.5\columnwidth]{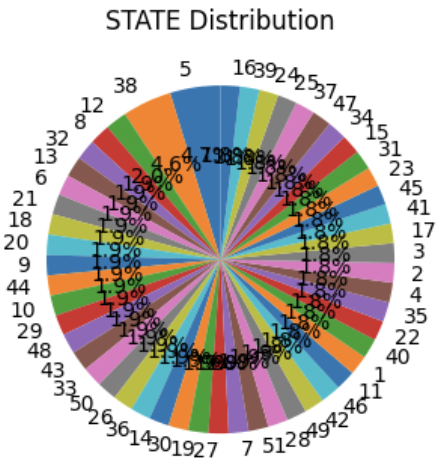}
        \label{fig:states}
    \end{center}
    \vskip -0.1in
\end{figure}

States were overall evenly represented in both datasets (not accounting for different relative sizes of state populations), although California ($5$) and Oregon ($38$) had higher relative representation (\autoref{fig:states}).

\begin{figure}[!htb]
    \vskip 0.1in
    \begin{center}
        \caption{Covariate imbalance in primary dataset.}
        \includegraphics[width=0.9\columnwidth]{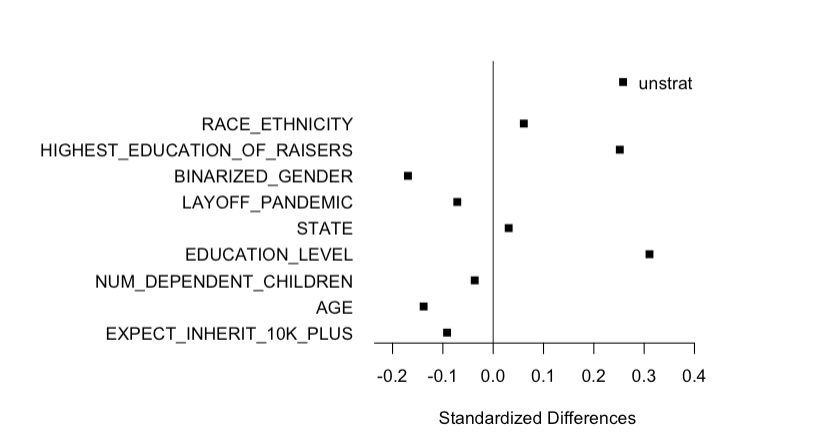}
        \label{fig:covbal}
    \end{center}
    \vskip -0.1in
\end{figure}

Imbalance between the treatment and control group covariates in the primary dataset is visualized as shown in \autoref{fig:covbal} using \texttt{xBalance} from the RITools R package \cite{RITools}. We see that without any manipulation, covariates are poorly balanced between the groups.

\subsection{Feature Engineering the Outcome Variable}\label{sec:feature_eng}

\begin{figure}[!ht]
    \vskip 0.1in
    \begin{center}
        \caption{Distribution of financial health outcomes in primary dataset.}
        \includegraphics[width=0.6\columnwidth]{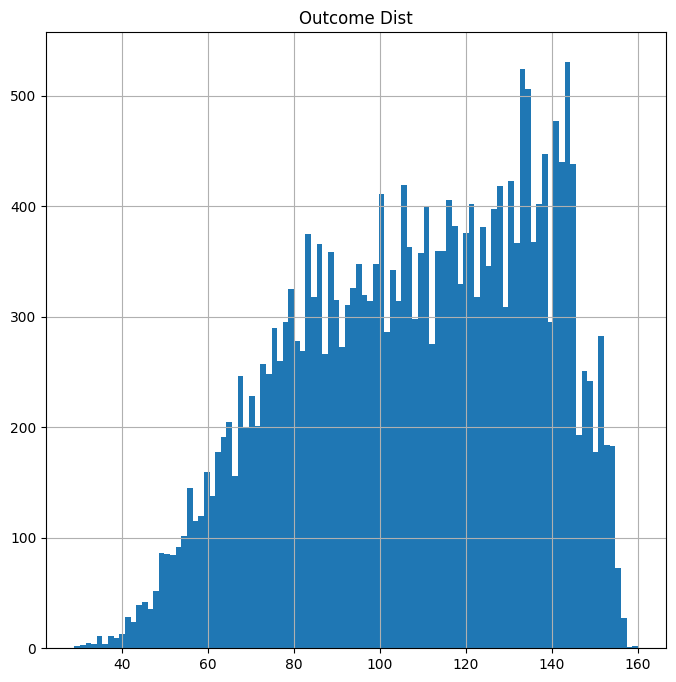}
        \label{fig:finhealth_dist}
    \end{center}
    \vskip -0.1in
\end{figure}

Towards having a single score indicating an individual's overall financial health, we engineer a \texttt{FIN\_HEALTH} variable by taking the sum of our financial health marker variables, after they have been standardized to have answer ranges of $0-9$ (\autoref{sec:health_markers}). For our primary analysis, this sum of financial health markers is our outcome variable. We chose not to perform further manipulation on our primary analysis to minimize potential for biases and assumptions in our analysis. Given markers $m_1, ..., m_{17}$ we calculate \textbf{financial health score} by: 
\begin{align*}
    \text{FIN\_HEALTH} = \text{sum}(m_1, ..., m_17)
\end{align*}
The resulting financial health score distribution for the primary dataset is as shown in \autoref{fig:finhealth_dist}. The distribution of financial health scores for our secondary analysis of high-school literacy courses can be found in \autoref{sec:secondary_finhealth_dist}.

Though we do not add marker coefficients in the primary analysis, we do acknowledge that each of the financial health markers is not equally important. As part of our sensitivity analysis (see \autoref{sec:scaled}), we scale each marker by our understanding of each marker's relative importance in determining an individual's financial health. Now given each marker $m_1, ..., m_{17}$ we calculate the \textbf{scaled financial health score} by:

\begin{align*}
    \text{FIN\_HEALTH\_SC} = \hspace{1mm} & m_1 + m_2 + m_3 + 0.5 \cdot m_4 + 0.5 \cdot m_5 \\
    & + m_6 + 0.25 \cdot m_7 + 0.25 \cdot m_8 \\
    & + 0.5 \cdot m_9 + m_{10} \\
    & + 0.75 \cdot m_{11} + m_{12} \\
    & + 1.25 \cdot m_{13} + 0.5 \cdot m_{14} \\
    & + 2 \cdot m_{15} + m_{16} + 0.5 \cdot m_{17}
\end{align*}

The distributions of scaled financial health score can be found in \autoref{sec:scaled_health_dists}.

\subsection{Data Limitations}

As all data in this study is purely observational, we conducted our experiments in the framework of observational study design. Furthermore, the data did not contain any information about the level of financial education received by study participants. As a result, our treatment variable $Z$ merely indicated whether a participant received education (treated) or not (control) and our treatment effect measured only the effect of attendance, regardless of level of engagement. As mentioned earlier, the NFCS dataset had very poor standardization of answer options, so we had to re-scale and re-order answer results for consistency. Notably, we did not consider the psychology or probability distributions for different ranges of answer options (eg. Does a $1-10$ answer range yield more moderate responses/less extreme responses compared to a $1-4$ answer range?).

\section{Methods}\label{sec:methods}

Since we had many pre-treatment covariates that could potentially confound our outcome, we elected to use causal inference methodologies that robustly identify treatment effects when pre-treatment covariates are present. In order to do this, we made some key assumptions. For each unit $i$, we have treatment indicator $Z_i$, pre-treatment covariates $X_i$, and potential outcomes $Y_i(0)$ and $Y_i(1)$, all assumed to be iid. This assumption was necessary due to the observational nature of our data - whereas in a randomization model with good controlled study design we could make the assumption that $Z \perp \{Y(0), Y(1)\}$, in the observational setting we need to make the assumption of \textbf{ignorability}:
$$ (Y(0), Y(1)) \perp Z | X $$
This assumption means that, given identical covariates $X$, exposure to the treatment is independent of the potential outcomes $Y(0)$ and $Y(1)$. By making this assumption, methods that effectively control for covariates $X$ can isolate observed differences in the outcome $Y$ due solely to the treatment. 


A quantity of interest in many of the estimators used in this section is the propensity score. The propensity score is  defined as the probability of receiving the treatment, conditioned on covariates and potential outcomes:

\begin{align*}
    e(X, Y(1), Y(0)) & = p(Z = 1|X, Y(0), Y(1)) \\
    & = p(Z=1|X) \\
    e(X) & = p(Z=1|X) 
\end{align*}
where the second equality comes from the assumption of ignorability. It can be shown that:
\begin{align*}
    & Z \perp (Y(1), Y(0)) | X \\
    \implies & Z \perp (Y(1), Y(0)) | e(X)
\end{align*}

The proof of this can be found in \autoref{app:proof1}. From this result, it follows that we can adjust our estimator using the single-dimensional propensity score instead of the potentially multi-dimensional covariates, and obtain equivalent results. In practice, logistic regression models are used to approximate the propensity score $\hat{e}(X)$. We follow this literature standard in our analysis.


\subsection{Generalized Lin's / Machine Learning Estimator}

Generalized Lin's estimator is one of several useful methods for identifying treatment effects in data with pre-treatment covariates. We note here that Generalized Lin's is typically employed only in the context of experimental data — despite this, Generalized Lin's provides a useful signal as to whether we ought to trust our estimates from AIPW, whose variance is reduced using techniques similar to Generalized Lin's. If the covariate distribution varies widely between treated and untreated groups, the mathematical justification of Generalized Lin's breaks down and we should thus be wary also of AIPW confidence intervals.

For generalized Lin's estimator, we estimate our average treatment effect (ATE) by building prediction models on treated units and control units, then using these prediction models to generate a hypothetical `complete` table of science whereby we know the outcome given treatment or control for each unit in the study. That is, for outcome variable $Y$, covariates $X$, and treatment $Z$, we learn a model $\hat{\mu_1}(\cdot)$ which models $Y \sim X$ for the subset of data where $Z=1$. Similarly, we learn a model $\hat{\mu_0}(\cdot)$ which models $Y \sim X$ for the subset of data where $Z=0$. By building these prediction models, we are inherently learning the effect of our covariates on the outcome variable. When we calculate our ATE in the end, each treatment unit now has a predicted `hypothetical control` outcome value to compare against, and vice versa for control units. Thus, we are able to better calculate the effect of the treatment in isolation from the effects of covariates.

When using modern machine-learning techniques it is critical to shift our models so that they are unbiased \footnote{With OLS models, the sum of residuals is $0$, so bias correction is unnecessary}. We adjust each $\hat{\mu_k}$ as follows:
\begin{align*}
    \tilde{\mu_k}(X) = \hat{\mu_k}(X) + \frac{1}{n_k}\sum_{i=1}^n Z_i \left( Y_i - \hat{\mu_k}(X_i) \right)
\end{align*}
We then calculate our ATE by:
\begin{align*}
    \hat{\tau}_{\text{pred}} & = \frac{1}{n} \left( \sum_{i: Z_i=1}Y_i + \sum_{i: Z_i=0}\tilde{\mu_1} (X_i) \right) \\
    & -\frac{1}{n} \left( \sum_{i: Z_i=1}\tilde{\mu_0}(X_i) + \sum_{i: Z_i=0} Y_i \right)
\end{align*}
Our variance estimation can then be calculated:
\begin{align*}
    \hat{V}_{\text{pred}} & = \frac{1}{n_1} \hat{\sigma}^2(1) + \frac{1}{n_0} \hat{\sigma}^2(0) + \frac{1}{n} \hat{\sigma}^2(\tau) \\
    \text{where }
    \hat{\sigma}^2(1) & = \frac{1}{n_1-1} \sum_{i: Z_i=1} (Y_i - \tilde{\mu_1}(X_i))^2 \\
    \text{and }
    \hat{\sigma}^2(0)& = \frac{1}{n_0-1} \sum_{i: Z_i=0} (Y_i - \tilde{\mu_0}(X_i))^2 \\
    \text{and }
    \hat{\sigma}^2(\tau) & = \frac{1}{n-1} \sum_{i=1}^n (\tilde{\mu}_1(X_i) - \tilde{\mu}_0(X_i) - \overline{\tilde{\mu}}_1 + \overline{\tilde{\mu}}_0)^2 \\
    \text{where }
    \overline{\tilde{\mu}}_1 & = \frac{1}{n} \sum_{j=1}^n \tilde{\mu}_1(X_j) \text{ and } \overline{\tilde{\mu}}_0  = \frac{1}{n} \sum_{j=1}^n \tilde{\mu}_0(X_j)
\end{align*}
Notably, in actual implementation, we perform \textbf{cross-fitting} to ensure that we do not overfit our models and that we get valid variance estimates. To perform cross-fitting, we split our data into two halves $I_1$ and $I_2$. We train a treated and control model for each half, then shift these models using the opposite half's data (to make them unbiased). With the unbiased models we calculate $\hat{\tau}_{\text{pred}^{I_1}}$ and $\hat{\tau}_{\text{pred}^{I_2}}$ for each half. Finally, we take the weighted average of these two estimators (weighting by relative size of each half in case of an imperfect split), to calculate our overall $\hat{\tau}_{\text{pred}}$. The full algorithm for cross-fitting with Generalized Lin's Estimator can be found in \autoref{sec:appendcrossfit}. 

For this analysis, we implemented cross-fitting in R with random forest as our model framework for training each $\hat{\mu_k}(\cdot)$ (models trained using the \texttt{randomForest} R package \cite{randomForest}).

\subsection{Propensity Matching}
Propensity matching is a matched pairs design technique for estimating the average treatment effect by comparing treated units with control units that have similar or identical covariates $X$. In this analysis, we used 1:1 matching, where control units were matched with at most one treated unit. In general, it is difficult to find exact matching for each control unit, and so approximate matching tecniques are used. In approximate matching, treatment units $i$ are matched with control units $m(i)$ where:
$$ m(i) = \arg \min_{k:Z_k = 0} d(X_i, X_k)$$
where $d$ is some distance metric in the covariate space. Commonly used is the Mahalanobis distance, defined as:
$$ d(X_i, X_k) = (X_i - X_k)^T \hat{\Sigma}^{-1} (X_i - X_k)$$
where $\hat{\Sigma}^{-1}$ is the sample covariance matrix of $X$. Similar is the robust Mahalanobis distance, which follows the same equation, but where $X$ is replaced by $\text{rank}(X)$.
In order to ensure units are close in propensity score, caliper matching further enforces the condition that:
$$ |\hat{e}(X_i) - \hat{e}(X_k)| \leq c \cdot sd(\hat{e}(X))$$
It is important to note that even after caliper matching, there may still be covariate imbalances between treatment and control groups — however, with effective matching, this is minimized, reducing the impact of confounders on biases on the treatment effect estimate. In 1:1 matching, we can first calculate an initial estimate for the treatment effect, and then adjust for any biases. The initial estimate is given by:
$$ \hat{\tau}^{m}= \frac{1}{n_1} \sum_{i:Z_i = 1} (Y_i(1) - Y_{m(i)}(0))$$
where $n_1$ is the number of treated units, and $Y_{m(i)}(0)$ is the observed outcome of the control unit matched with treatment unit $i$. The bias in this term can be corrected for using the following \cite{abadie2011bias}:
$$ \hat{B} = \frac{1}{n_1} \sum_{i:Z_i = 1} \hat{B}_i$$
$$ \hat{B}_i = \hat{\mu}_0(X_i) - \hat{\mu}_0(X_{m(i)})$$
where $\hat{\mu}_0$ is an estimate of $E[Y|Z=0, X = X]$ which can be fit using linear regression methods. The bias adjusted estimator is then given by:
$$ \hat{\tau}^{caliper} = \hat{\tau}^m - \hat{B}$$
In this analysis, we use the \texttt{DOS2}, \texttt{optmatch}, and \texttt{rcbalance} packages in R to perform 1:1 matching using robust Mahalanobis distance, with different caliper values of 0.1, 0.2, and 0.05, before using these matched pairs to compute a bias-corrected estimate of the average treatment effect \cite{DOS2, optmatch, rcbalance}.

\subsection{IPW}
Inverse propensity score weighting (IPW) is the basis for two estimators of the treatment effect used in the present analysis — the Horvitz-Thompson estimator \cite{horvitz1952generalization}, and the Hajek estimator. Both estimators rely on the following result, which holds under ignorability. The proof is shown in \autoref{app:proof2}:
\begin{align*}
    E[Y(1)] & =  E\left[\frac{ZY}{e(X)}\right] \\
    E[Y(0)] & =  E\left[\frac{(1 - Z)Y}{1-e(X)}\right] 
\end{align*}
This result motivates an estimator:
$$ \hat{\tau}^{ht} = \frac{1}{n} \sum_{i=1}^n \frac{Z_iY_i}{\hat{e}(X_i)} - \frac{(1 - Z_i)Y_i}{(1 - \hat{e}(X_i))}$$
Essentially, this estimator is constructed by weighting the observation of each individual in the sample by their propensity score. Since $e(x)$ is bounded to be between 0 and 1, this estimator can often experience instability for observations that have high or low propensity scores - to remedy this issue, truncation is sometimes used, where propensity scores are limited as follows:
$$ \hat{e}(X) \leftarrow \min(0.975, \max(\hat{e}(X), 0.025)) $$

An additional shortcoming of this estimator is that it is not invariant under transformations of the outcome variable. To show this, let $\tilde{Y} = Y + c$:
\begin{align*}
    \hat{\tau}^{ht}(Y) - \hat{\tau}^{ht}(\tilde{Y}) & = \frac{1}{n}\sum_{i=1}^n \frac{Z_i Y_i}{\hat{e}(X_i)}- \frac{(1 - Z_i)Y_i}{(1 - \hat{e}(X_i))} \\
    & - \frac{1}{n}\sum_{i=1}^n \frac{Z_i (Y_i+c)}{\hat{e}(X_i)}- \frac{(1 - Z_i)(Y_i+c)}{(1 - \hat{e}(X_i))}
\end{align*}
If we rearrange terms to have a common denominator and factor them out, this can equivalently be written as:
\begin{align*}
    \frac{1}{n} \sum_{i=1}^n \frac{1}{\hat{e}(X_i)(1 - \hat{e}(X_i))} & \bigg(Z_iY_i - Z_i Y_i \hat{e}(X_i) - Y_i \hat{e}(X_i) \\
    & + Z_i Y_i \hat{e}(X_i) Z_i Y_i + Z_i Y_i\hat{e}(X_i) \\
    & - Z_i c + Z_ic \hat{e}(X_i) + Y_i \hat{e}(X_i) + \\
    & c\hat{e}(X_i) - Z_iY_i \hat{e}(X_i) - Z_i c \hat{e}(X_i)\bigg)
\end{align*}
When canceling terms, we are left with:
$$ \frac{1}{n} \sum_{i=1}^n \frac{c( \hat{e}(X_i) - Z_i)}{\hat{e}(X_i)(1 -\hat{e}(X_i))} \neq 0 $$
concluding the proof. In this analysis, we have constructed our outcome variable as a composition of several variables, and so this property is undesirable. This motivates the location-invariant Hajek estimator:
$$ \hat{\tau}^{hajek} = \frac{\sum_{i=1}^n \frac{Z_i Y_i}{\hat{e}(X_i)}}{\sum_{i=1}^n \frac{Z_i}{\hat{e}(X_i)} } - \frac{\sum_{i=1}^n \frac{(1 - Z_i Y_i}{(1 - \hat{e}(X_i))}}{\sum_{i=1}^n \frac{(1 - Z_i)}{(1 - \hat{e}(X_i))} }$$

For the implementation of both of these methods, the propensity score model $\hat{e}(X_i)$ was fit using logistic regression. The Hajek estimator was calculated using the \texttt{PSweight} package \cite{psweight}, and the Horvitz-Thompson estimator was implemented directly using base R functions.

\subsection{AIPW}

A potential issue with using IPW based estimators is that they have high variance. This motivates the Augmented Inverse Propensity Score Weighted estimator (AIPW), which applies the idea of inverse propensity weighting to a general function $g$ of the covariates $X$. Directly applying such a function in the IPW introduces bias into the estimator, but this bias term can be computationally corrected for, resulting in an overall unbiased estimator. To show this, we can observe that for any function $g(X)$, the following is true:
\begin{align*}
    E\left[ \frac{1}{n} \sum_{i=1}^n \frac{Z_i g(X_i)}{e(X_i)}  - \frac{1}{n} \sum_{i=1}^n g(X_i) \right] \\
    = \frac{1}{n} \sum_{i=1}^n \bigg( E\bigg[\frac{Z_ig(X_i)}{e(X_i)}\bigg] - E[g(X_i)] \bigg)\\
    = \frac{1}{n} \sum_{i=1}^n \bigg( E\bigg[\frac{E[Z_i|X_i]g(X_i)}{e(X_i)} \bigg] - E[g(X_i)] \bigg) \\
    = \frac{1}{n} \sum_{i=1}^n \bigg( E\bigg[\frac{e(X_i)g(X_i)}{e(X_i)} \bigg] - E[g(X_i)] \bigg)\\
    = \frac{1}{n} \sum_{i=1}^n ( E[g(X_i)] - E[g(X_i)] ) = 0
\end{align*}
Similarly, it can be shown that:
\begin{align*}
    E\left[ \frac{1}{n} \sum_{i=1}^n \frac{(1 - Z_i) g(X_i)}{(1 - e(X_i))}  - \frac{1}{n} \sum_{i=1}^n g(X_i) \right] = 0
\end{align*}
We can therefore form a reduced estimator by first fitting a logistic regression model to estimate $\hat{e}(X)$, fitting linear models $\hat{\mu}_1(X)$ and $\hat{\mu}_0(X)$ that fit:
\begin{align*}
    \mu_1(X) = E[Y|Z = 1, X = x] \\
    \mu_0(X) = E[Y|Z = 0, X = x]
\end{align*}
and fitting adjusted linear models:
\begin{align*}
    \hat{\mu}_1^{adj} & = \frac{1}{n} \sum_{i=1}^n \frac{Z_i(Y_i - \hat{\mu}_1(X_i))}{\hat{e}(X_i)} + \hat{\mu}_1(X_i) \\
    \hat{\mu}_0^{adj} & = \frac{1}{n} \sum_{i=1}^n \frac{(1 - Z_i)(Y_i - \hat{\mu}_0(X_i))}{(1 - \hat{e}(X_i))} + \hat{\mu}_0(X_i)
\end{align*}
From this, we form the following AIPW estimator:
$$ \hat{\tau}^{aipw} = \hat{\mu}_1^{adj} - \hat{\mu}_1^{adj}$$

In this analysis, this estimator was calculated using the \texttt{AIPW} package \cite{aipw_lib}, which forms confidence intervals using cross-fitting to estimate the variance. This estimator is known as the `doubly-robust' estimator, because it only requires either the outcome models $\hat{\mu}^{adj}$ or the propensity scores $e(X)$ to be accurate in expectation in order for the treatment effect estimate to be correct \cite{bang2005doubly}. Thus, AIPW incorporates robustness benefits from both Generalized Lin's Estimator and IPW.

\section{Primary Experiments and Results}\label{sec:experiments} 

Estimates of the average treatment effect of financial education on financial health score for all estimation techniques are shown in \autoref{tab:fin_all_unscaled}. Also included are variance estimates and $95$\% confidence intervals for the treatment effect.

\begin{table}[h]
  \caption{ATEs for primary analysis.}
  \label{tab:fin_all_unscaled}
  \centering
  \begin{tabular}{|c|ccc|}
    \hline
    \textbf{Estimator} & \textbf{ATE} & \textbf{$\hat{V}$} & \textbf{95\% CI} \\
    \hline
    Generalized Lin's & 3.99 & 0.13 & 3.27 - 4.70 \\
    \hline
    Horvitz-Thompson & 3.71 & 0.35 & 3.02 - 4.40 \\
    \hline
    Hajek & 3.71 & 0.25 & 3.31 - 4.18 \\
    \hline
    AIPW & 3.55 & 0.382 & 2.80 - 4.30 \\
    \hline
    Matching (caliper=0.1) & 2.91 & 0.20 & 2.04 - 3.78 \\
    \hline
    Matching (caliper=0.2) & 2.75 & 0.20 & 1.88 - 3.61 \\
    \hline
    Matching (caliper=0.05)  & 3.13 & 0.20 & 2.26 - 4.00 \\
    \hline
  \end{tabular}
\end{table}

\section{Secondary Experiments and Results: HS Financial Literacy Courses}

Estimates of the average treatment effect of high-school based financial literacy education on financial health score for all estimation techniques are shown in \autoref{tab:fin_hs_unscaled}. Also included are variance estimates and $95$\% confidence intervals for the treatment effect.
\begin{table}[H]
  \caption{ATEs for unscaled secondary analysis.}
  \label{tab:fin_hs_unscaled}
  \centering
  \begin{tabular}{|c|ccc|}
    \hline
    \textbf{Estimator} & \textbf{ATE} & \textbf{$\hat{V}$} & \textbf{95\% CI} \\
    \hline
    Generalized Lin's & 1.53 & 0.43 & 0.24 - 2.81 \\
    \hline
    Horvitz-Thompson & 2.28 & 0.72 & 0.87 - 3.69 \\
    \hline
    Hajek & 2.28 & 0.76 & 0.79 - 3.46 \\
    \hline
    AIPW & 1.85 & 0.69 & 0.51 - 3.19 \\
    \hline
    Matching (caliper=0.1) & 1.48 & 0.49 & 0.10 - 2.85 \\
    \hline
    Matching (caliper=0.05)  & 1.44 & 0.49 & 0.06 - 2.81 \\
    \hline
    Matching (caliper=0.2)  & 1.58 & 0.49 & 0.21 - 2.96 \\
    \hline
  \end{tabular}
\end{table}

\section{Sensitivity Analysis}

\subsection{Modified Financial Health Score Function}\label{sec:scaled}
In order to confirm the robustness of our results, the average treatment effects were also estimated for \texttt{FIN\_HEALTH\_SC} outcomes with different weights applied to each financial health marker, as described at the end of \autoref{sec:feature_eng}. These average treatment effects, and their corresponding variance estimates and confidence intervals can be found in  \autoref{tab:fin_all_scaled} and \autoref{tab:fin_hs_scaled}.

\subsubsection{Primary Analysis}
\begin{table}[H]
  \caption{ATEs for scaled primary analysis.}
  \label{tab:fin_all_scaled}
  \centering
  \begin{tabular}{|c|ccc|}
    \hline
    \textbf{Estimator} & \textbf{ATE} & \textbf{$\hat{V}$} & \textbf{95\% CI} \\
    \hline
    Generalized Lin's & 2.70 & 0.10 & 2.09 - 3.31 \\
    \hline
    Horvitz-Thompson & 2.59 & 0.29 & 2.02 - 3.17 \\
    \hline
    Hajek & 2.59 & 0.33 & 1.84 - 3.12 \\
    \hline
    AIPW & 2.47 & 0.32 & 1.83 - 3.10 \\
    \hline
    Matching (caliper=0.1) & 2.03 & 0.14 & 1.29 - 2.78 \\
    \hline
    Matching (caliper=0.2) & 1.91 & 0.14 & 1.17 - 2.66 \\
    \hline
    Matching (caliper=0.05)  & 2.21 & 0.14 & 1.47 - 2.95 \\
    \hline
  \end{tabular}
\end{table}
\subsubsection{Secondary Analysis}

See \autoref{tab:fin_hs_scaled} for the weighted financial health score function results on the HS financial education analysis.

\begin{table}
  \caption{ATEs for scaled secondary analysis.}
  \label{tab:fin_hs_scaled}
  \centering
  \begin{tabular}{|c|ccc|}
    \hline
    \textbf{Estimator} & \textbf{ATE} & \textbf{$\hat{V}$} & \textbf{95\% CI} \\
    \hline
    Generalized Lin's & 1.34 & 0.32 & 0.24 - 2.44 \\
    \hline
    Horvitz-Thompson & 1.82 & 0.60 & 0.63 - 3.00 \\
    \hline
    Hajek & 1.82 & 0.55 & 0.90 - 3.07 \\
    \hline
    AIPW & 1.46 & 0.58 & 0.32 - 2.61 \\
    \hline
    Matching (caliper=0.1) & 1.11 & 0.36 & -0.08 - 2.29 \\
    \hline
    Matching (caliper=0.05)  & 1.06 & 0.36 & -0.12 - 2.24 \\
    \hline
    Matching (caliper=0.2)  & 1.19 & 0.36 & 0.01 - 2.37 \\
    \hline
  \end{tabular}
\end{table}

\subsection{Testing Different Calipers}

We conducted further sensitivity analysis by repeating the 1:1 matched pairs design with several different calipers. While a 0.1 caliper is commonly used, and seemed the most apt for our estimation due to its relatively strong resultant covariate balance and average difference between propensity scores in each pair, we also estimated our ATE with calipers of 0.05 and 0.2 to test for robustness. We observed estimates for the ATE and Variance that were very similar across each caliper.

\section{Discussion}\label{sec:discussion}

\subsection{Effect of Financial Literacy Education on Financial Health}

All estimation methods in \autoref{tab:fin_all_unscaled} provide strong statistical evidence to indicate that financial education positively impacts financial health. All of these results were significant at the 5\% confidence level. As expected, the inverse propensity-weighted estimators had the highest variance, but still provided sufficient evidence to reject the null hypothesis that financial education has no impact on financial health.

Notably, however, the magnitude of the effect was quite small, ranging from $1.17-3.31$ points from minimum to max across all methods. Relative to the scale of the financial health score used here (average score was $112$ for treated units, and $115$ for controls), our results indicate an isolated treatment effect of approximately \textbf{3\%} improvement due to receiving financial literacy education.

Propensity matching was highly effective at reducing covariate imbalance between matched pairs in the setting of the primary treatment variable. As seen in Appendix \autoref{fig:cov_imb_all}, standardized differences within the unstratified data were relatively high, most notably for \texttt{highest\_education\_of\_raisers}, \texttt{education\_level} and our propensity scores, at $0.25$, $0.31$, and $0.4$, respectively. But within the matched pairs, each difference dropped to $< 0.025$.

\subsection{Effect of High-School Literacy Education on Financial Health}

We observe a weaker average treatment effect when specifically investigating the effects of financial education received in high-school on financial health score (\autoref{tab:fin_hs_unscaled}).
Although weaker, the results do again provide statistical evidence to reject the null at the 5\% confidence level. Notably, the control group for this secondary analysis includes individuals who may have received financial education elsewhere from high-school, weakening our power to identify an effect.

Propensity matching effectively reduced covariate imbalance between matched pairs in our secondary experiment as well, albeit not to the same degree as the primary analysis. This may be due to a smaller pool of controls for selecting optimal matches \autoref{sec:tc_dists}. As we show in Appendix \autoref{fig:cov_imb_hs}, standardized differences within the unstratified data were relatively high, most notably with \texttt{education\_level} exceeding 0.4, and our propensity scores approaching 0.6. After matching, differences dramatically improve, with a maximum distance of around $0.25$.

\subsection{Sensitivity Analysis}
For the primary analysis, average treatment effect estimates using our weighted financial health score outcome function were broadly similar to those in the original analysis (\autoref{tab:fin_all_scaled}). Both the direction and the significance of the results remained consistent with the original analysis, providing evidence that our analysis is robust to variations in the exact computation of the financial health outcome.

When focusing on HS-based financial education, the average treatment effect estimates using the weighted financial health score were not all significant at the 5\% level (\autoref{tab:fin_hs_scaled}). In particular, caliper matching with caliper sizes of $0.1$ and $0.05$ yielded $95$\% confidence intervals containing $0$. This is an indicator that the matched pairs estimator was not robust to scaling in the outcome variable. However, all other estimators still indicated a positive treatment effect at the 5\% significance level, indicating that on the whole, there is evidence that HS-based financial education has a positive effect on financial health, even when manipulating the exact financial health measurement function.

\subsection{Limitations}
While the NFCS data was sprawling in both number of respondents and number of topics covered, it had some limitations for our question of interest. The data did not describe how long ago survey respondents received financial literacy education, making it harder to filter the data and isolate the effect for people who have recently received the treatment, but with enough time for its potential benefits to be realized. Many variables were binned into groupings that limited the amount of information available to researchers. Perhaps most importantly, many responses we used in calculating financial health outcome scores were self-assessments of the individual's financial health, and these standards may vary from person to person.

\section{Conclusion}\label{sec:conclusion}

This study used data from the 2021 NFCS to understand the causal effect of financial education on aggregated financial health outcomes. Our results provide strong statistical evidence that financial education positively impacts financial health scores, though the magnitude of that effect is potentially quite weak. These results were robust to numerous causal effect estimation techniques, as well as sensitivity analysis on matching caliper choices and differently constructed composite financial health scores. Our results may prove useful to policymakers considering implementing financial education requirements in public school systems. Moreover, these findings suggest that any individual considering financial education is highly likely to benefit from such an education. \\

Future works should focus on devising a more informed formulation for the financial health score, based on domain expertise. None of the authors of this work are primarily involved in the study of finance, education, or long term financial health trajectories. An extension of the current research complemented by domain-expert insight in the outcome variable would likely yield much more interpretable results for estimating the real-world benefit of the ATE.

\clearpage

\section{Contributions}

Arnav and Charles performed initial literature review. All contributors collaborated in determining experimental process, and defining covariates, treatment variables, and outcomes. Daniel wrote the data processing code and Generalized Lin's Estimator experiment code. Charles wrote the propensity matching experiment code. Arnav wrote the IPW and AIPW experiment code. All contributors contributed in writing the final paper.

\section{Code}

All code for this work can be found at \url{https://github.com/danielfrees/finlitCausal}.


\bibliography{finlit}
\bibliographystyle{icml2023}

\newpage
\onecolumn
\appendix

\section{Covariate List}\label{sec:cov_list}

\begin{table}[!ht]
  \centering
  \caption{List of Covariates \vspace{2mm}}
  \label{tab:covariates}
  \begin{tabular}{|p{\dimexpr\linewidth-2\tabcolsep}|}
    \hline
    \textbf{Covariates} \\
    \hline
    \small
    \begin{enumerate}
      \item RACE\_ETHNICITY
      \item EDUCATION\_LEVEL
      \item HIGHEST\_EDUCATION\_OF\_RAISERS
      \item NUM\_DEPENDENT\_CHILDREN
      \item BINARIZED\_GENDER
      \item AGE
      \item LAYOFF\_PANDEMIC
      \item EXPECT\_INHERIT\_10K\_PLUS
      \item STATE
    \end{enumerate} \\
    \hline
  \end{tabular}
\end{table}

\section{Financial Health Markers}\label{sec:health_markers}

\begin{table}[!ht]
  \centering
  \caption{List of Financial Health Markers \vspace{2mm}}
  \label{tab:financial_markers}
  \begin{tabular}{|p{\dimexpr\linewidth-2\tabcolsep}|}
    \hline
    \textbf{Financial Health Markers} \\
    \hline
    \small
    \begin{enumerate}
      \item 'SATISFACTION\_WITH\_FINANCIAL\_CONDITION'
      \item 'SPENDING\_COMPARISON\_TO\_INCOME'
      \item 'DIFFICULTY\_COVERING\_EXPENSES'
      \item 'EMERGENCY\_FUNDS'
      \item 'CONFIDENCE\_GET\_2000'
      \item 'CREDIT\_RECORD\_RATING'
      \item 'CHECKING\_ACCOUNT'
      \item 'SAVINGS\_ACCOUNT'
      \item 'OVERDRAW\_CHECKING\_ACCOUNT'
      \item 'REGULAR\_CONTRIBUTION\_TO\_RETIREMENT'
      \item 'OTHER\_INVESTMENTS'
      \item 'ALWAYS\_PAY\_CR\_FULL\_12MO'
      \item 'USED\_PAYDAY\_LOAN'
      \item 'DEBT\_COLLECTED\_12MO'
      \item 'TOO\_MUCH\_DEBT\_STRENGTH'
      \item 'D2D\_FINANCIAL\_SKILL'
      \item 'FINANCIAL\_KNOWLEDGE\_ASSESS'
    \end{enumerate} \\
    \hline
  \end{tabular} \\
\end{table}

\section{Conditional Independence based on Propensity}\label{app:proof1}
Our goal is to show:
\begin{align*}
    Z \perp (Y(1), Y(0)) | X \implies & Z \perp (Y(1), Y(0)) | e(X)
\end{align*}
Equivalently, we can show:
\begin{align*}
    P(Z=1|Y(1), Y(0), e(X)) & = p(Z=1|e(X)) \\
\end{align*}
For the top term, we can argue:
\begin{align*}
    P(Z=1|Y(1), Y(0), e(X)) & = E[Z|Y(1), Y(0), e(x)] \\
    & =  E[E[Z|Y(1),Y(0), X]|Y(1, Y(0), e(X)]\\
    & =  E[e(X)|Y(1), Y(0), e(X)] \\
    & =  e(X)
\end{align*}
Similarly, for the bottom term:
\begin{align*}
    P(Z=1|e(X)) = E[Z|e(X)] & = E[E[Z|X]|e(X)] \\
    & = E[e(X)|e(X)] \\
    & = e(X)
\end{align*}

\section{Propensity estimator is equal in expectation to treatment effect}\label{app:proof2}
We will show the proof for:
$$ E[Y(1)] = E\left[\frac{ZY}{e(X)}\right]$$
as the proof for the other expression is analogous:
\begin{align*}
    E\left[\frac{ZY}{e(X)}\right] & =  E[E\left[\frac{ZY}{e(X)}|X\right]] \\
    &=  E[\frac{1}{e(X)}E[ZY(1)|X]] \\
    & =  E[\frac{1}{e(X)} E[Z|X]E[Y(1)|X]] \\
    & =  E[\frac{1}{e(X)} e(X) E[Y(1)|X]] \\
    & =  E[E[Y(1)|X]] \\
    & =  E[Y(1)]
\end{align*}
where the move from $Y$ to $Y(1)$ comes from the consistency of the outcome depending on the treatment, and the third equality comes from the conditional independence of $Z$ and $Y(1)$. 

\clearpage 

\section{Cross-fitting Algorithm for Generalized Lin's / Machine Learning Estimator}\label{sec:appendcrossfit}

Here we describe the algorithm for calculating the Generalized Lin's / Machine Learning average treatment effect estimate and variance estimate via cross-fitting on outcome $Y$, covariates $X = X_1, ... X_n$, and treatment $Z \in {0,1}$ using randomForest as the model. Note that we denote two halves of the data used for cross-fitting as $I_1, I_2$, and in half $k$ we denote the treated examples as $I_{k}^+$ and controls as $I_{1}^-$.\\

\begin{algorithm*}[htbp]
   \caption{ML Estimator Crossfitting ATE and Variance Estimation}
   \label{alg:crossfit}
\begin{algorithmic}
   \STATE {\bfseries Input:} data $D$, size $n$ \vspace{3mm}
   \STATE $I_1, I_2 \leftarrow$ random split of  $D$ into halves. \vspace{3mm}
  \FOR{$k \in {1,2}$} \vspace{2mm}
     \STATE $\hat{\mu_1}^k(X) \leftarrow \text{randomForest}(Y\sim X_1 + ... + X_n, \text{data}=I_k, \text{subset } Z=1)$ \vspace{2mm}
     \STATE $\hat{\mu_0}^k(X) \leftarrow \text{randomForest}(Y\sim X_1 + ... + X_n, \text{data}=I_k, \text{subset } Z=0)$ \vspace{2mm}
     \STATE $\tilde{\mu_1}^k(\cdot) = \hat{\mu_1}^k(\cdot) + \frac{1}{|I_{(1-k)}^+|} \sum_{i\in I_{(1-k)}}Z^{(i)}(Y^{(i)}-\hat{\mu_1}^k(X^{(i)}))$ \vspace{2mm}
     \STATE $\tilde{\mu_0}^k(\cdot) = \hat{\mu_0}^k(\cdot) + \frac{1}{|I_{(1-k)}^-|} \sum_{i\in I_{(1-k)}}(1-Z^{(i)})(Y^{(i)}-\hat{\mu_0}^k(X^{(i)}))$ \vspace{2mm}
     \STATE $\hat{\tau}^{I_k} \leftarrow \frac{1}{|I_k|} \left( \sum_{i \in I_k} Z_iY_i + (1-Z_i)\tilde{\mu_1}^{(1-k)}(X_i) - \sum_{i \in I_k} (1-Z_i)Y_i + Z_i\tilde{\mu_0}^{(1-k)}(X_i) \right) $
  \ENDFOR \vspace{3mm}
  \STATE $\hat{\tau}^{\text{pred}} \leftarrow \frac{|I_1|}{n}\hat{\tau}^{I_1} + \frac{|I_2|}{n}\hat{\tau}^{I_2}$ \vspace{2mm}
  \FOR{$k \in {1,2}$} \vspace{2mm}
    \STATE $\hat{\sigma}^2_{I_k}(1) \leftarrow \frac{1}{|I_k^+|-1}\sum_{i\in I_k}Z_i(Y_i - \tilde{\mu}_1^{I_{(1-k)}}(X_i))^2$ \vspace{2mm}
    \STATE $\hat{\sigma}^2_{I_k}(0) \leftarrow \frac{1}{|I_k^-|-1}\sum_{i\in I_k}(1-Z_i)(Y_i - \tilde{\mu}_0^{I_{(1-k)}}(X_i))^2$ \vspace{2mm}
    \STATE $\bar{\tilde{\mu}}_1^{I_{(1-k)}} \leftarrow \frac{1}{|I_k|} \tilde{\mu}_1^{I_{(1-k)}}(X_i)$ \vspace{2mm}
    \STATE $\bar{\tilde{\mu}}_0^{I_{(1-k)}} \leftarrow \frac{1}{|I_k|} \tilde{\mu}_0^{I_{(1-k)}}(X_i)$ \vspace{2mm}
    \STATE $\hat{\sigma}^2_{I_k}(\tau) \leftarrow \frac{1}{|I_k|-1}\sum_{i\in I_k} \left( \tilde{\mu}_1^{I_{(1-k)}}(X_i)) - \tilde{\mu}_0^{I_{(1-k)}}(X_i) - \left( \bar{\tilde{\mu}}_1^{I_{(1-k)}} - \bar{\tilde{\mu}}_0^{I_{(1-k)}}\right) \right)^2$ \vspace{2mm}
    \STATE $\hat{V}^{I_k} \leftarrow \frac{1}{|I_k+|}\hat{\sigma}^2_{I_k}(1)+ \frac{1}{|I_k-|}\hat{\sigma}^2_{I_k}(0) + \frac{1}{|I_k|}\hat{\sigma}^2_{I_k}(\tau))$ \vspace{2mm}
  \ENDFOR \vspace{3mm}
  \STATE $\hat{V} \leftarrow \left( \frac{|I_1|}{n} \right)^2\hat{V}^{I_1} + \left( \frac{|I_2|}{n} \right)^2\hat{V}^{I_2}$ \vspace{2mm} 
  \STATE \textbf{return}  $\hat{\tau}^{\text{pred}}, \hat{V}$
  
\end{algorithmic}
\end{algorithm*}

\clearpage 
\section{Complete Covariate Distributions}\label{sec:cov_dists}

\begin{figure}[htb]
    \vskip 0.1in
    \begin{center}
        \caption{Distribution of covariates in primary dataset.}
        \includegraphics[width=0.9\columnwidth]{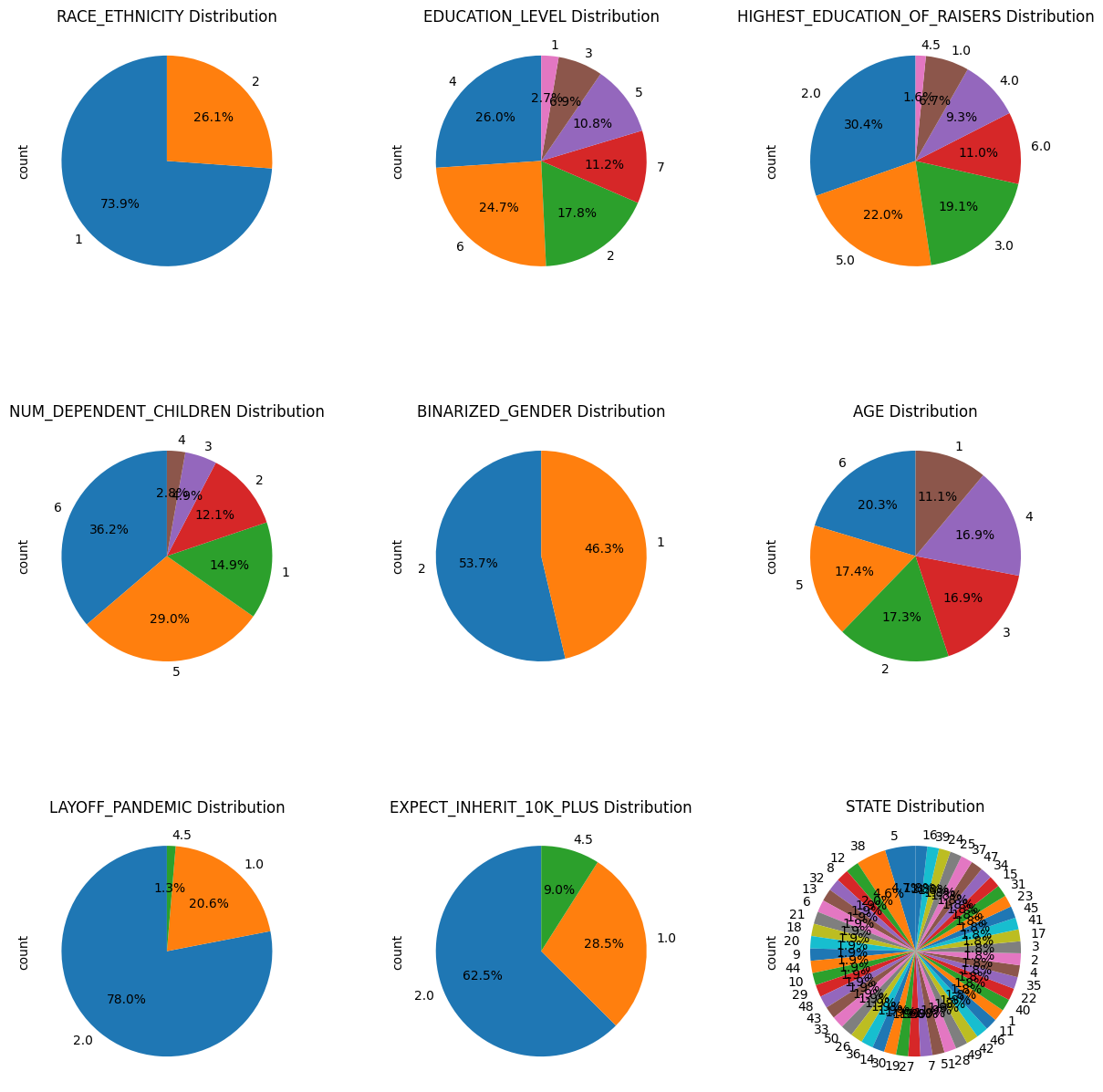}
        \label{fig:primary_covs}
    \end{center}
    \vskip -0.1in
\end{figure}

\begin{figure}[htb]
    \vskip 0.1in
    \begin{center}
        \caption{Distribution of covariates in HS literacy dataset.}
        \includegraphics[width=0.9\columnwidth]{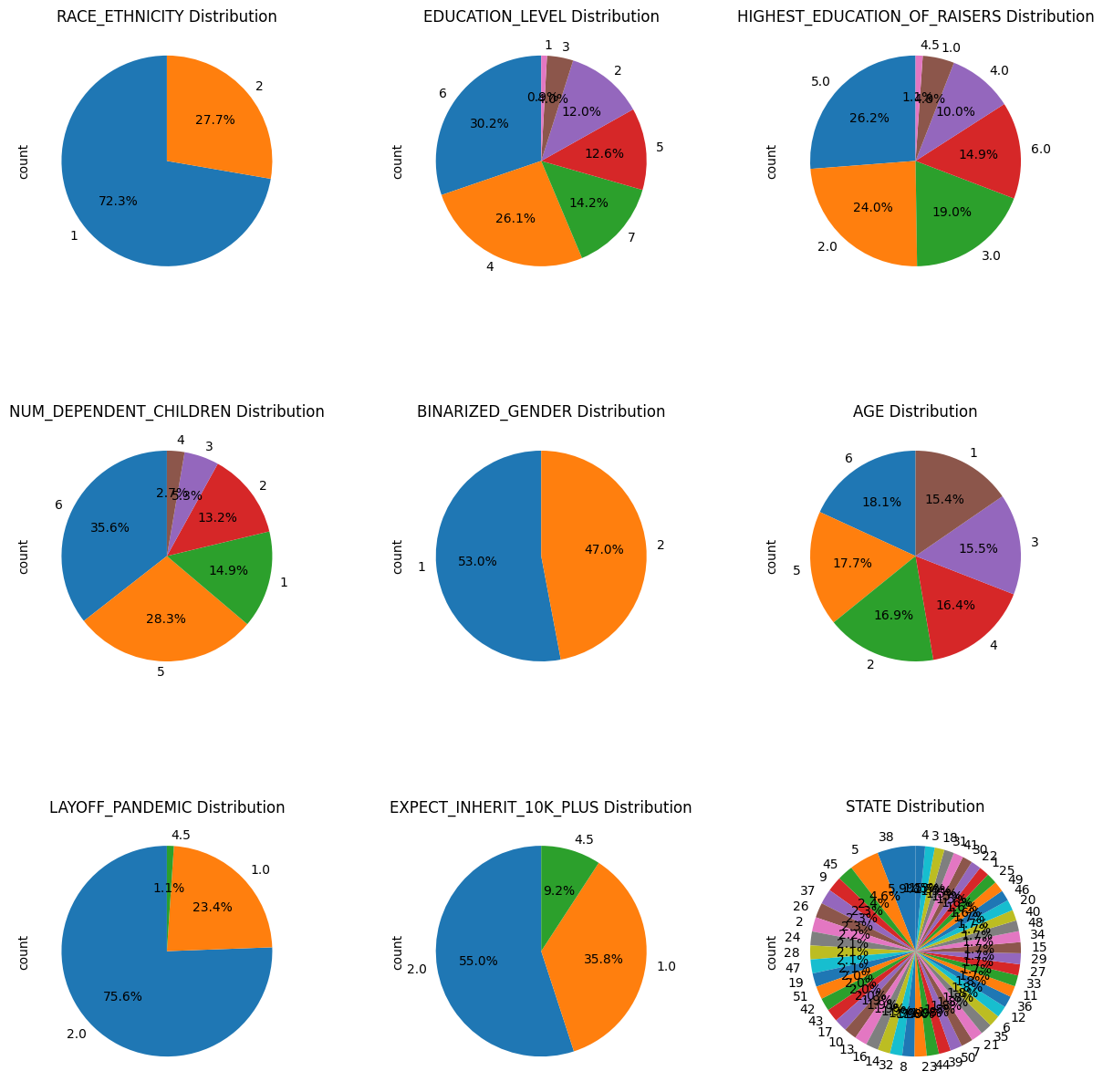}
        \label{fig:hs_covs}
    \end{center}
    \vskip -0.1in
\end{figure}

\begin{figure}[!ht]
    \vskip 0.1in
    \begin{center}
        \caption{Covariate imbalance before and after matched pairs stratification in primary dataset.}
        \includegraphics[width=1\columnwidth]{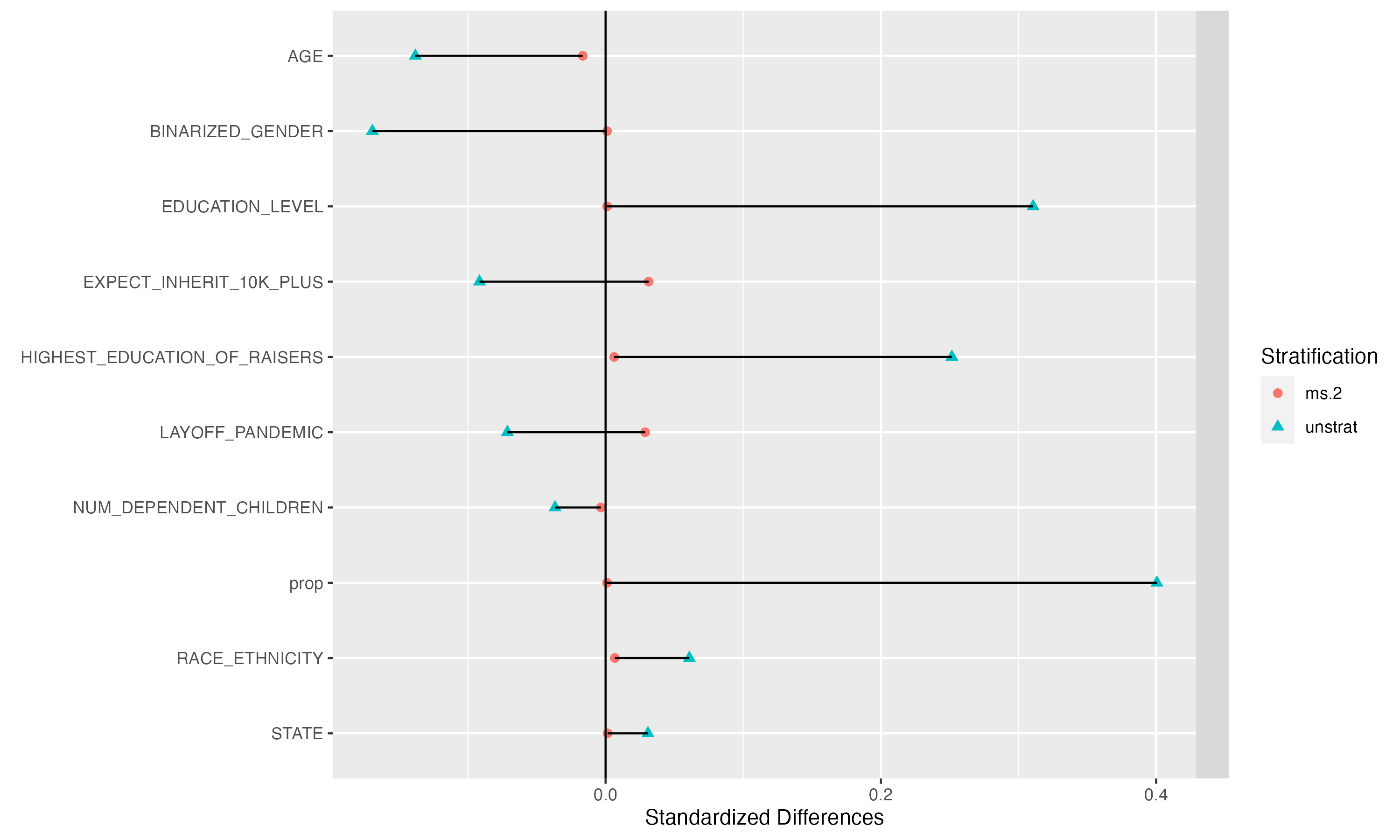}
        \label{fig:cov_imb_all}
    \end{center}
    \vskip -0.1in
\end{figure}

\begin{figure}[!ht]
    \vskip 0.1in
    \begin{center}
        \caption{Covariate imbalance before and after matched pairs stratification in HS literacy dataset.}
        \includegraphics[width=1\columnwidth]{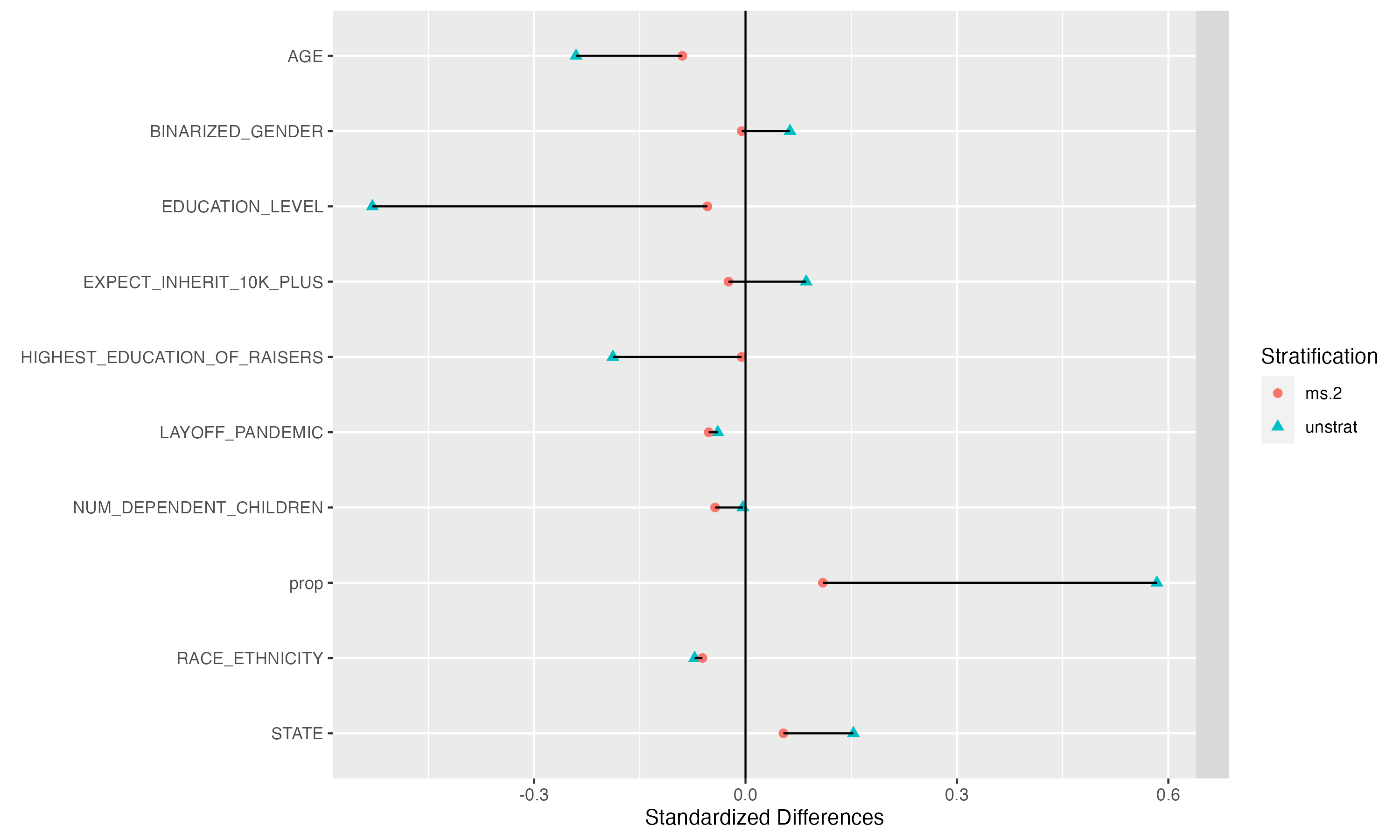}
        \label{fig:cov_imb_hs}
    \end{center}
    \vskip -0.1in
\end{figure}

\clearpage 

\section{Treatment and Control Distributions}\label{sec:tc_dists}

\begin{figure}[htb]
    \vskip 0.1in
    \begin{center}
        \caption{Distribution of treatment in primary dataset.}
        \includegraphics[width=0.42\columnwidth]{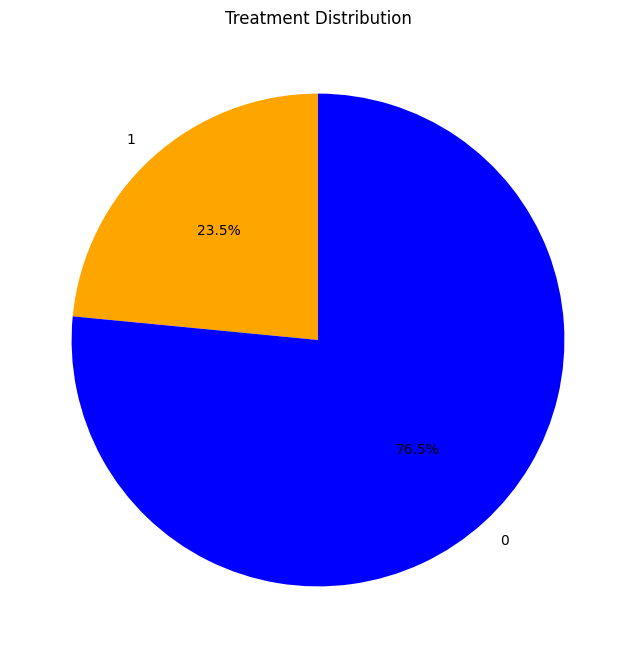}
        \label{fig:primary_tc}
    \end{center}
    \vskip -0.1in
\end{figure}

\begin{figure}[htb]
    \vskip 0.1in
    \begin{center}
        \caption{Distribution of treatment in HS literacy dataset.}
        \includegraphics[width=0.42\columnwidth]{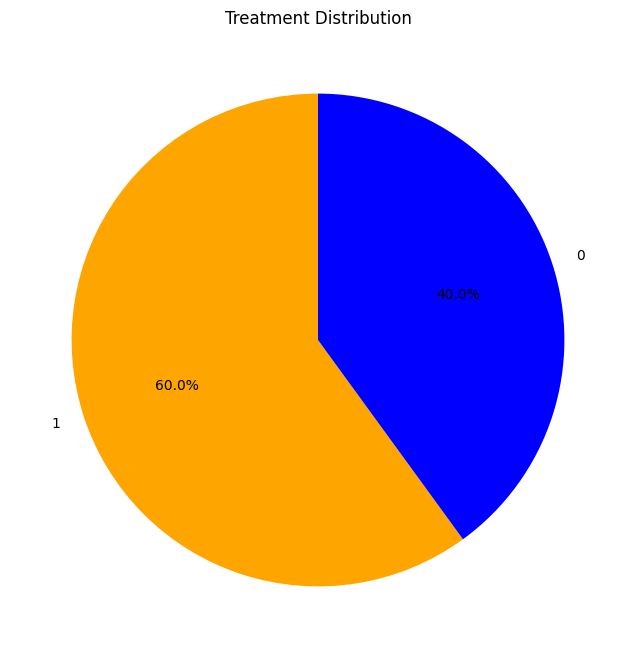}
        \label{fig:hs_tc}
    \end{center}
    \vskip -0.1in
\end{figure}

\clearpage

\section{Secondary Analysis Financial Health Distribution}\label{sec:secondary_finhealth_dist}

\begin{figure}[htb]
    \vskip 0.1in
    \begin{center}
        \caption{Distribution of financial health outcome in HS literacy dataset.}
        \includegraphics[width=0.9\columnwidth]{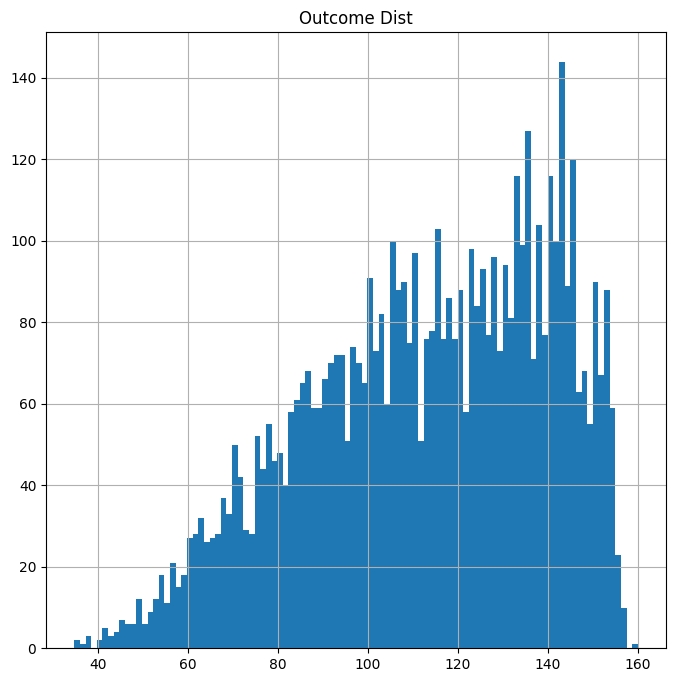}
        \label{fig:hs_health_dist}
    \end{center}
    \vskip -0.1in
\end{figure}

\clearpage

\section{Scaled Financial Health Distributions}\label{sec:scaled_health_dists}

\begin{figure}[htb]
    \vskip 0.1in
    \begin{center}
        \caption{Distribution of scaled financial health outcome in primary dataset.}
        \includegraphics[width=0.9\columnwidth]{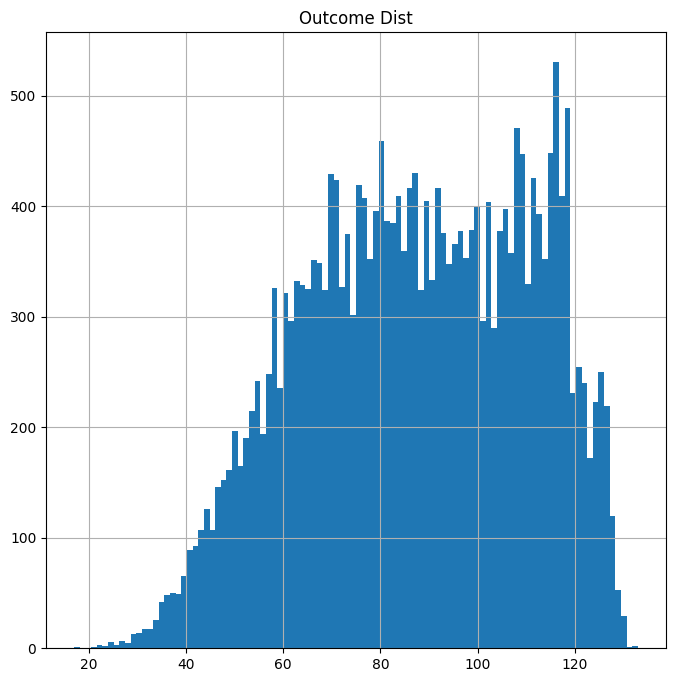}
        \label{fig:primary_scaled_health_dist}
    \end{center}
    \vskip -0.1in
\end{figure}

\begin{figure}[htb]
    \vskip 0.1in
    \begin{center}
        \caption{Distribution of scaled financial health outcome in HS literacy dataset.}
        \includegraphics[width=0.9\textwidth]{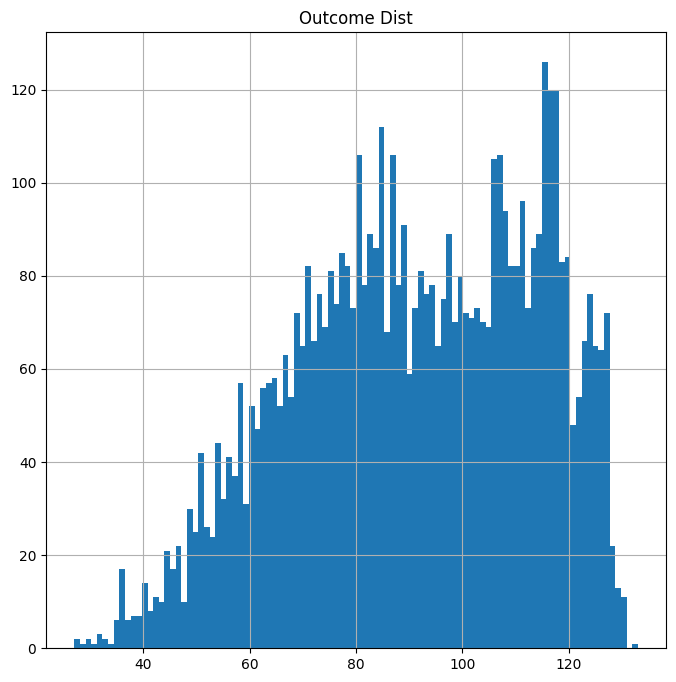}
        \label{fig:hs_scaled_health_dist}
    \end{center}
    \vskip -0.1in
\end{figure}


\end{document}